\documentclass[aps,prl,twocolumn,superscriptaddress]{revtex4-1}
\usepackage{graphicx}
\usepackage{dcolumn}
\usepackage{bm}
\usepackage{amsmath}
\usepackage{float}
\usepackage[colorlinks=true,citecolor=blue,linkcolor=magenta]{hyperref}
\allowdisplaybreaks

\begin{document}

\title{Multiphoton interference in quantum Fourier transform circuits and applications to quantum metrology}

\author{Zu-En Su}
\author{Yuan Li}
\affiliation{Hefei National Laboratory for Physical Sciences at Microscale and Department of Modern Physics,\\
	University of Science and Technology of China, Hefei, Anhui 230026, China}
\affiliation{CAS Centre for Excellence and Synergetic Innovation Centre in Quantum Information and Quantum Physics,\\
	University of Science and Technology of China, Hefei, Anhui 230026, China}

\author{Peter P. Rohde}
\affiliation{\mbox{Centre for Quantum Software \& Information (QSI), Faculty of Engineering \& Information Technology,}\\
	University of Technology Sydney, Sydney, NSW 2007, Australia}

\author{He-Liang Huang}
\author{Xi-Lin Wang}
\author{Li Li}
\author{Nai-Le Liu}
\affiliation{Hefei National Laboratory for Physical Sciences at Microscale and Department of Modern Physics,\\
	University of Science and Technology of China, Hefei, Anhui 230026, China}
\affiliation{CAS Centre for Excellence and Synergetic Innovation Centre in Quantum Information and Quantum Physics,\\
	University of Science and Technology of China, Hefei, Anhui 230026, China}

\author{Jonathan P. Dowling}
\affiliation{Hearne Institute for Theoretical Physics and Department of Physics \& Astronomy, \\
	Louisiana State University,	Baton Rouge, Louisiana 70803, USA}

\author{Chao-Yang Lu}
\author{Jian-Wei Pan}
\affiliation{Hefei National Laboratory for Physical Sciences at Microscale and Department of Modern Physics,\\
	University of Science and Technology of China, Hefei, Anhui 230026, China}
\affiliation{CAS Centre for Excellence and Synergetic Innovation Centre in Quantum Information and Quantum Physics,\\
	University of Science and Technology of China, Hefei, Anhui 230026, China}

\date{\today}

\begin{abstract}
Quantum Fourier transforms (QFT) have gained increased attention with the rise of quantum walks, boson sampling, and quantum metrology. Here we present and demonstrate a general technique that simplifies the construction of QFT interferometers using both path and polarization modes. On that basis, we first observed the generalized Hong-Ou-Mandel effect with up to four photons. Furthermore, we directly exploited number-path entanglement generated in these QFT interferometers and demonstrated optical phase supersensitivities deterministically.
\end{abstract}

\maketitle

Quantum interference lies at the heart of quantum mechanics. Increasing the number of single photons and the complexity of optical circuits are key advances for a quantum advantage in many photonic quantum processing tasks \cite{pan2012}; including quantum computing \cite{steane1996,*knill2001,*kok2007}, quantum simulation \cite{aspuru2012,*georgescu2014}, and quantum metrology \cite{lee2002,*giovannetti2011}.

The Hong-Ou-Mandel (HOM) effect \cite{hong1987} is regarded as one of the quintessential quantum interference phenomena. In the original experiment, two identical single photons interfered and bunched in a two-mode quantum Fourier transform (QFT) interferometer (i.e., a balanced beam splitter). Generally, $n$ identical single photons interfering in an $n$-mode QFT interferometer \cite{marek1997,*vourdas2005,*lim2005a} will lead to a higher-dimensional bunching effect \cite{tichy2010,*spagnolo2013,rigovacca2016}, which is expected to play an important role in understanding and exploiting multiphoton interference. A recent application of the QFT is to use it for stringent and efficient assessment of boson sampling \cite{tichy2014a}, which can guarantee the results contain genuine quantum interference \cite{tichy2014b}. The QFT interferometer has been constructed on chip with up to eight modes \cite{crespi2016}. However, only three-photon assessment was demonstrated \cite{carolan2015},  due in part to the relatively high loss of the optical circuit. For this scheme to work with more photons, it is essential to construct large-scale and low-loss QFT interferometers.

\begin{figure}[b]
\includegraphics[width=3.4 in]{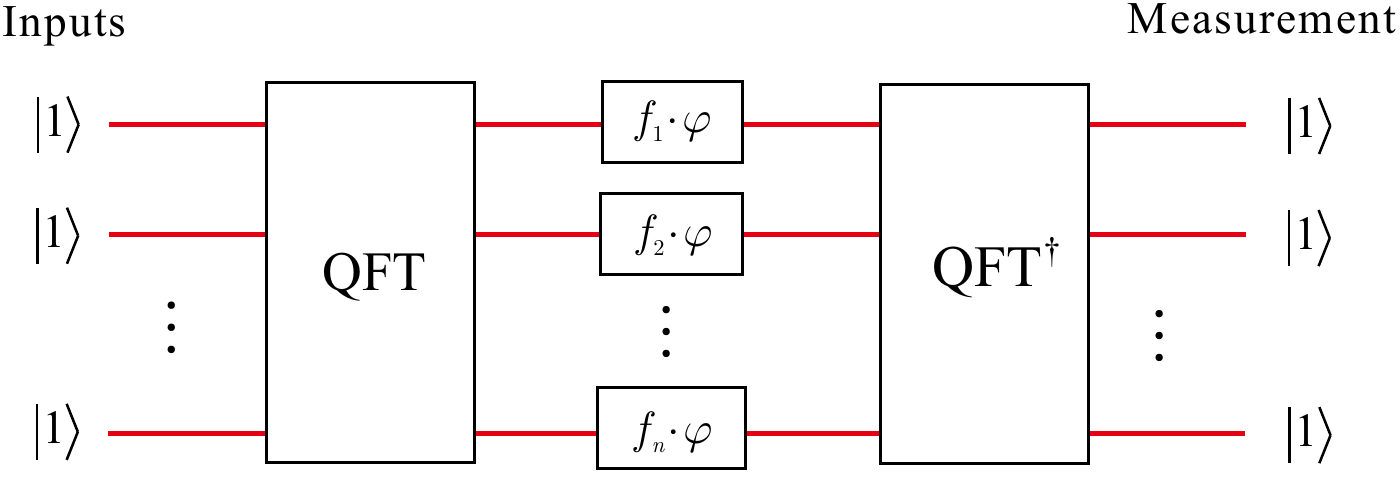}
\caption{\label{fig:scheme} Quantum metrology scheme of the QFT interferometers using single-photon inputs. The QFT acts as the number-path entanglement generator, while the inverse transform $\mbox{QFT}^{\dag}$ is used for un-entangling the probe. Counting coincidence events with one photon per output mode leads to the probability distribution that is used to estimate the unknown phase $\varphi$.}
\end{figure}

Quantum metrology is another important application intimately related to quantum interference. One of the most versatile quantum metrology devices---the Mach-Zehnder interferometer (MZI), is made up of two balanced beam splitters. Naturally, $m$-mode QFT interferometers were proposed to construct multimode MZIs for precision improvement \cite{weihs1996, chaboyer2015} or simultaneous estimation of multiple phases \cite{humphreys2013,ciampini2016,*szczykulska2016}. Recently, Motes and Olson \textit{et al.} \cite{motes2015,olson2016linear} pointed out that an $n$-mode MZI fed with a single photon into each arm can be used to beat the shot noise limit (SNL) deterministically [see Fig.~\ref{fig:scheme}], requiring neither nonlinear nor probabilistic preparation of entanglement. However, since multimode MZIs consist of a QFT and an inverse QFT interferometer, having higher loss and lower stability than a single QFT interferometer, only one \cite{weihs1996} and two \cite{chaboyer2015} photons were tested in a three-mode MZI so far. It remains a challenge to observe multiphoton interference in multimode MZIs to beat the SNL.

\begin{figure*}[t]
	\includegraphics[width=5.5 in]{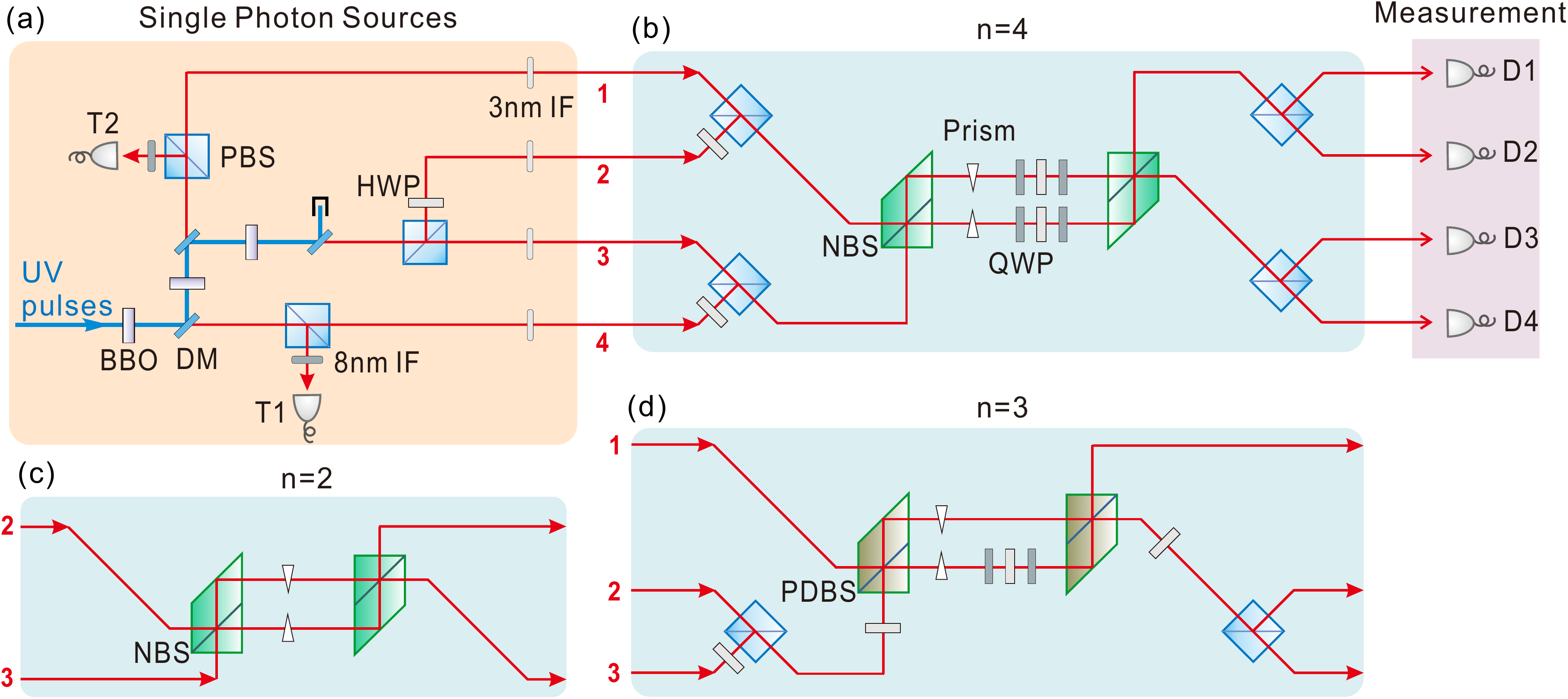}
	\caption{\label{fig:setup} Experimental setup. (a) The single photon sources. Photons are produced in three nonlinear crystals (BBO) via spontaneous parametric down conversion. Motorized translation stages $\Delta {d_1}$ -- $\Delta {d_3}$ (not drawn in the figure) were used to synchronize the delays among paths one to four. The quantum metrology optical circuit with (b) four, (c) two, and (d) three single-photon inputs. The labels are: DM, dichroic mirror; PBS, polarizing beam splitter; PDBS, polarization-dependent beam splitter; NBS, non-polarizing beam splitter; HWP (QWP), half (quarter) wave plate; Prism used as phase shifter between different paths; IF, interferential filter; D1, D2, D3, D4, T1, T2, fiber-coupled single-photon detectors.}
\end{figure*}

In this work, we develop a general approach to simplifying the construction of QFT interferometers using both path and polarization modes, which makes it possible to reduce resources as much as 75\% when compared to devices using only path modes. We report the first experimental demonstration of the generalized HOM effect with up to four photons. Moreover, we constructed multimode MZIs using two cascaded QFTs and observed phase supersensitivies deterministically.

The single-photon inputs were generated via three spontaneous parametric down conversion (SPDC) sources [Fig.~\ref{fig:setup}(a)], each emitting one pair of photons ${\left| H \right\rangle _s}{\left| V \right\rangle _i}$, where $H$ and $V$ denote horizontal and vertical polarizations, and $s$ and $i$ correspond to the signal and idler path modes, respectively. For $n=2$, one pair of SPDC photons was enough, while for $n=3$, another SPDC was added, and all three SPDCs were used for $n=3$. For the latter two cases, post selection of a four-fold (six-fold) coincidence, consisting of one (two) triggers, ensures that only three (four) photons enter the setup in separate modes with a negligible higher-order noise.

The QFT interferometers were constructed with low-loss bulk-optical elements. In order to decrease the number of beam splitters and improve the interference stability, we exploited polarization and path modes simultaneously. This simplification makes us able to construct the QFT interferometers with only one non-polarizing beam splitter (NBS) for $n=4$ and 2 [Fig.~\ref{fig:setup}(b), (c)], and one polarization-dependent beam splitter (PDBS) for $n=3$ [Fig.~\ref{fig:setup}(d)]. More details can be found in the Supplemental Material \cite{SUPPLEMENTALMATERIAL}.

\begin{figure*}[t]
\includegraphics[width=5.5 in]{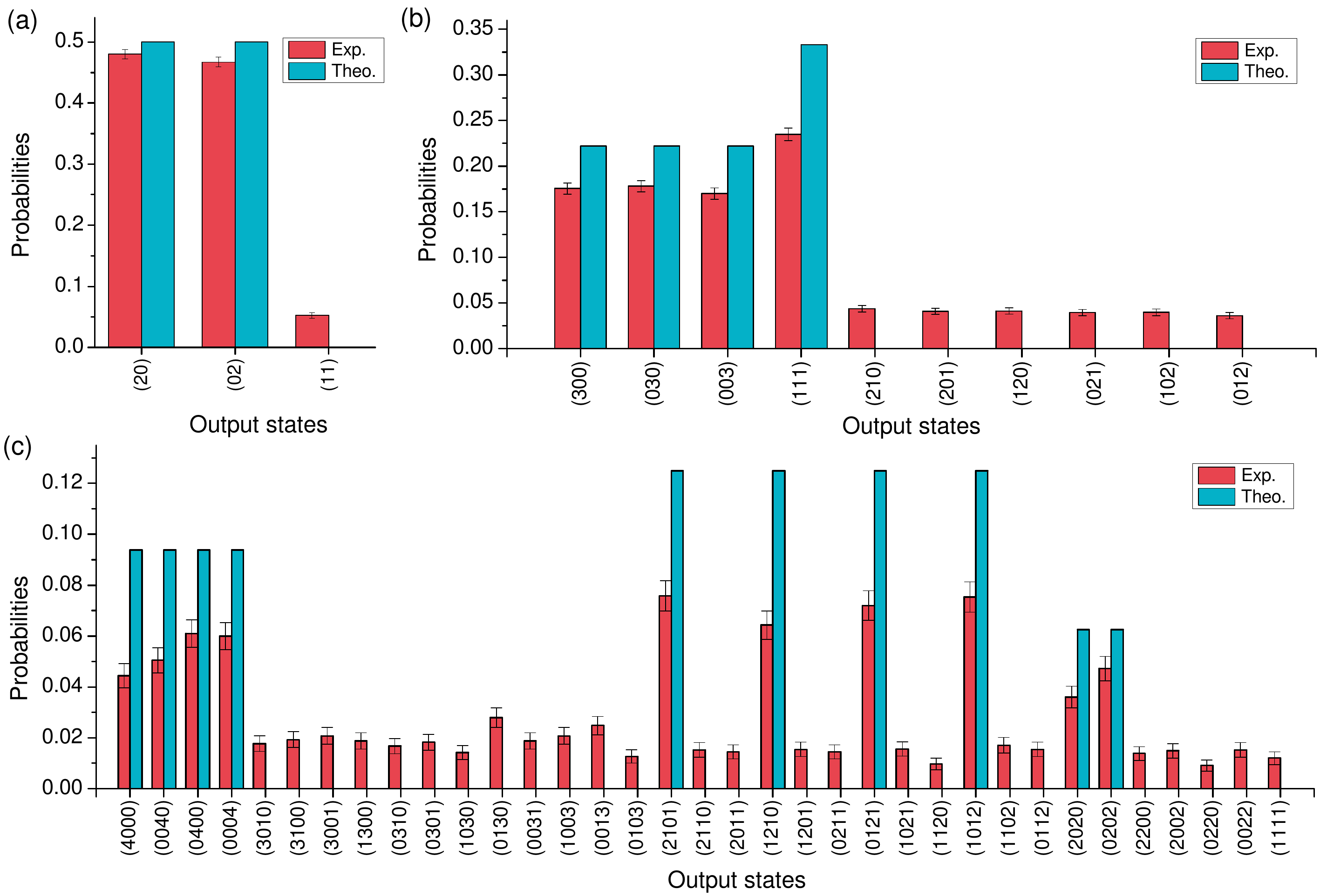}
\caption{\label{fig:qft} Experimental results of the generalized HOM effect. (a) $n=2$; (b) $n=3$; (c) $n=4$. These bunching output states with photon number $\ge 2$, such as (300) and (210), were measured by multiplexing the single-photon detectors with arrays of beam splitters. Error bars are one standard deviation due to propagated Poissonian statistics.}
\end{figure*}

{\em Generalized HOM effect.---} Going through the QFT interferometers, the single-photon inputs will evolve as,
\begin{alignat}{2}
&\left| {11} \right\rangle  \hspace{3pt} &\to \hspace{3pt}& (\left| {20} \right\rangle  - \left| {02} \right\rangle )/\sqrt 2, \label{eq:1}\\
&\left| {111} \right\rangle \hspace{3pt} &\to \hspace{3pt}& \frac{{\sqrt 2 }}{3}(\left| {300} \right\rangle  + \left| {030} \right\rangle  + \left| {003} \right\rangle ) - \frac{1}{{\sqrt 3 }}\left| {111} \right\rangle, \label{eq:2}\\
&\left| {1111} \right\rangle \hspace{3pt} &\to \hspace{3pt}& \frac{{\sqrt 6 }}{8}(\left| {4000} \right\rangle  - \left| {0400} \right\rangle  + \left| {0040} \right\rangle  - \left| {0004} \right\rangle ) \nonumber \\
& &\hspace{3pt} + \hspace{3pt}& \frac{{\sqrt 2 }}{4}(\left| {1210} \right\rangle  - \left| {2101} \right\rangle  + \left| {1012} \right\rangle  - \left| {0121} \right\rangle ) \nonumber\\
& &\hspace{3pt} - \hspace{3pt}& \frac{1}{4}(\left| {2020} \right\rangle  - \left| {0202} \right\rangle \label{eq:3}).
\end{alignat}
Other terms are destructively interfered to zero according to the so-called zero-transmission law \cite{tichy2010}, which predicts which output configurations will be strictly suppressed in the generalized HOM effect. The theoretical and experimental probability distributions are shown in Fig.~\ref{fig:qft}. We use the fidelity defined as $F = \sum\nolimits_i {\sqrt {{p_i}{q_i}} }$ to quantify the similarity between the experimental probability distribution $\{ {p_i}\} $ and the theoretical one $\{ {q_i}\}$, with respect to the three states in Eqs. (\ref{eq:1} -- \ref{eq:3}). We obtained fidelities of $0.973 \pm 0.001$, $ 0.871 \pm 0.004$, and $ 0.765 \pm 0.008$ for $n = 2$, 3, and 4, respectively. To distinguish an effect associated with classical particles, we calculated the experimental violation of Eqs. (\ref{eq:1} -- \ref{eq:3}) as ${\upsilon_n}  = {N_s}/{N_t}$; the ratio of the number of predicted suppressed events ${N_s}$ to the total number of events ${N_t}$. For $n = 2$, 3, and 4, the violations with indistinguishable single photons are $\upsilon _2^{{\mbox{ind}}} = 0.052 \pm 0.001$, $\upsilon _3^{{\mbox{ind}}} = 0.24 \pm 0.01$, $\upsilon _4^{{\mbox{ind}}} = 0.41 \pm 0.03$, compared with the larger values $\upsilon _2^{\mbox{d}} = 0.47 \pm 0.01$, $\upsilon _3^{\mbox{d}} = 0.68 \pm 0.08$, $\upsilon _4^{\mbox{d}} = 0.75 \pm 0.14$ with distinguishable single photons (coherent states). (See more details in the Supplemental Material \cite{SUPPLEMENTALMATERIAL}.)

The results can also be viewed as a nonclassicality witness of the input sources \cite{rigovacca2016}. We calculated the average second-order correlation function, defined as ${\overline G _n} = \frac{2}{{n(n - 1)}}\sum\nolimits_{i < j} {{n_i}{n_j}{p_{ij}}} $, where ${n_i}$ is the photon number in the $i$-th mode and ${p_{ij}}$ is the coincidence probability between the $i$-th and $j$-th modes. We obtained ${\overline G _2} = 0.052 \pm 0.005$, ${\overline G _3} = 0.396 \pm 0.007$ and ${\overline G _4} = 0.556 \pm 0.011$, significantly violating their corresponding classical lower bounds $(1 - 1/n)$, ${G_2} = 0.5$, ${G_3} = 0.67$, ${G_4} = 0.75$, and indicating good average pairwise indistinguishability of the input sources.

\begin{figure}[t]
\includegraphics[width=3.4 in]{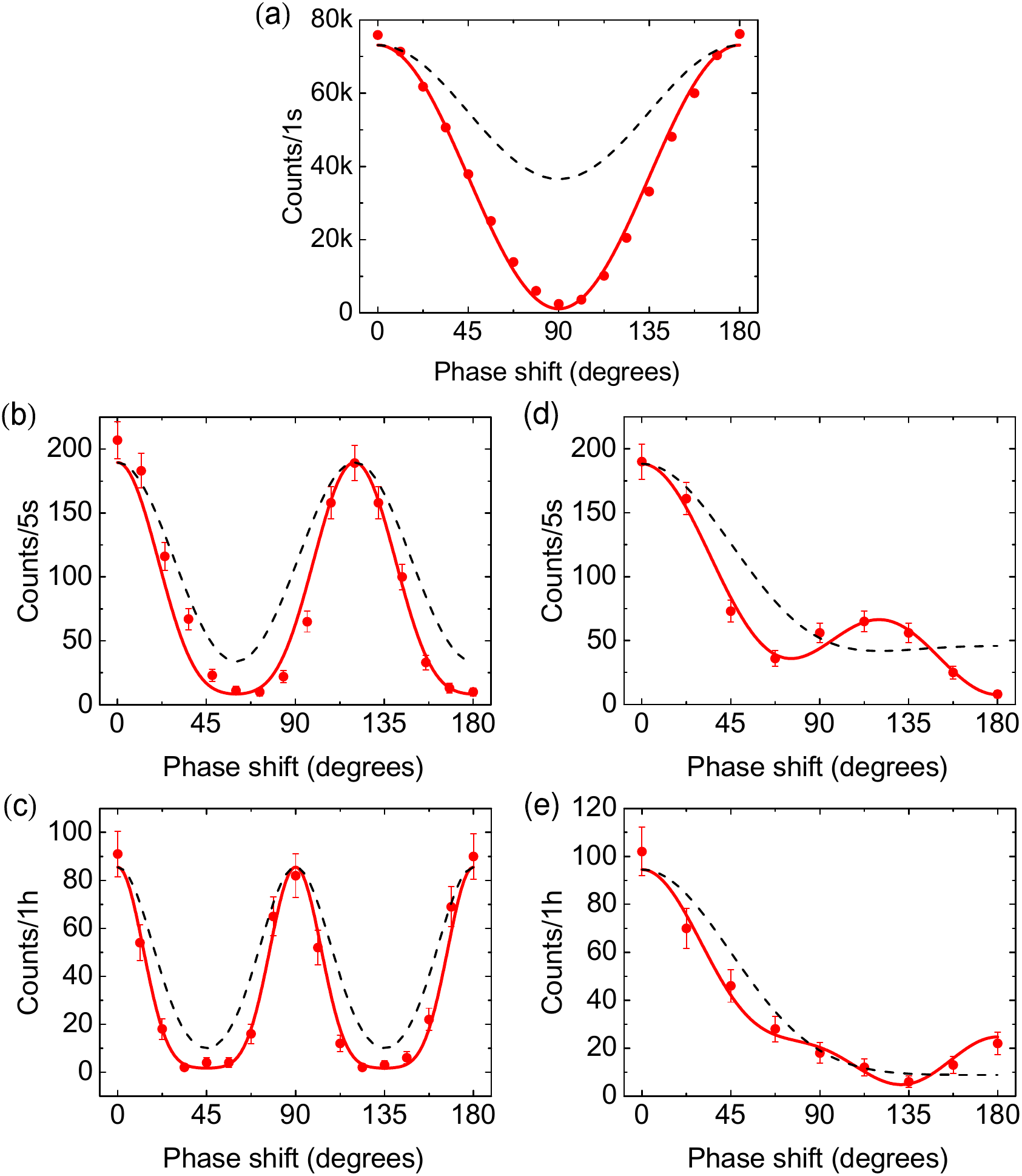}
\caption{\label{fig:fringe} Measuring counts as function of phase shift $\varphi$ for (a -- c) linear and (d, e) delta phase distribution. The fringes exhibit 1, 1.5 and 2 distinct oscillations within a half phase cycle for (a) $n=2$, (b, d) $n=3$ and (c, e) $n=4$ respectively. Error bars are one standard deviation due to propagated Poissonian statistics. The solid red line is a fit to the case of  indistinguishable single photons, while the dashed line is the limiting distribution of distinguishable single photons. From (a) to (e), the experimental (classical-limit) visibilities are $0.97 \pm 0.02$ ($0.50$), $0.908 \pm 0.020$ ($0.609$), $0.962 \pm 0.025$ ($0.790$), $0.927 \pm 0.037$ ($0.636$), $0.919 \pm 0.045$ ($0.829$) respectively. Here visibility is defined as $({\mbox{Counts}_{\max }} - {\mbox{Counts}_{\min }})/({\mbox{Counts}_{\max }} + {\mbox{Counts}_{\min }})$, different from the fitted parameter called effective visibility in the main text and Supplemental Material \cite{SUPPLEMENTALMATERIAL}.}
\end{figure}

{\em Quantum metrology based on the QFT.---} Our scheme has different phase distributions $\{ {f_j} \cdot \varphi \} _{j = 1}^n$, as illustrated in Fig.~\ref{fig:scheme}. To demonstrate the basic principle, we chose two phase distributions (linear phase $f_j^{{\mbox{lin}}} = j - 1$ and delta phase $f_j^\delta  = {\delta _{j,1}}$) and implemented two- to four-photon experiments. As shown in Fig.~\ref{fig:fringe}, all experimental fringes exhibit phase superresolution and, most importantly, oscillate with a better visibility than the corresponding classically limited distributions, which are given in the Supplemental Material \cite{SUPPLEMENTALMATERIAL}. Different from other schemes based on engineered entangled states, e.g., the N00N state \cite{boto2000,*dowling2008}, the QFT-based quantum metrology scheme directly exploits deterministically generated entanglement [Eqs. (\ref{eq:1} -- \ref{eq:3})]. Thus, we do not need to worry about post-selection efficiency when trying to demonstrate phase supersensitivity \cite{resch2007,*nagata2007}.

The linear phase scheme has phase sensitivity that scales as $O(1/{n^{3/2}})$ \cite{motes2015}. Unfortunately, the high sensitivity is due to the linearly increasing phase shift $\{ (j - 1)\varphi \} _{j = 1}^n$, but not the quantum nature of multiphoton interference \cite{d1997,*soderholm2003,*chwedenczuk2013}. Nevertheless, the linear scheme is superresolving and could have applications to quantum microscopy \cite{juffmann2016multi}. Note that the largest relative phase shift is $(n - 1)\varphi \sim n\varphi$. If we run a classical two-mode MZI $n$ times to measure the largest relative phase shift $n\varphi $, its classical sensitivity is $\Delta (n\varphi ) = O(1/{n^{1/2}})$, that means $\Delta (\varphi ) = O(1/{n^{3/2}})$, giving the same improvement as the linear phase scheme. In fact, it has been pointed out that Kitaev's phase-estimation algorithm, based on the QFT with a similar linearly increasing phase shift, cannot beat the SNL unless one uses adaptive measurements \cite{higgins2007,*giovannetti2006}.

\begin{table}
	\caption{\label{tab:table} Measuring phase sensitivity $\Delta\varphi$ against the number of photons $n$. Here HL denotes Heisenberg limit.}
	\begin{ruledtabular}
		\begin{tabular}{ccccc}
			\textrm{Photon-number ($n$)}&
			\textrm{${\Delta \varphi_{\mbox{ideal}}}$}&
			\textrm{${\Delta \varphi_{\mbox{exp}}}$}&
			\textrm{SNL}&
			\textrm{HL}\\
			\colrule
			2 & 0.500 & $0.515 \pm 0.013$ & 0.707 & 0.500\\
			3 & 0.433 & $0.491 \pm 0.015$ & 0.577 & 0.333\\
			4 & 0.408 & $0.458 \pm 0.027$ & 0.500 & 0.250\\
		\end{tabular}
	\end{ruledtabular}
\end{table}

The delta function proved to be the best phase distribution to demonstrate phase supersensitivity, although $\Delta \varphi $ only scales as $\sqrt {\frac{n}{{8(n - 1)}}}$ \cite{olson2016linear}. As the number of photons increases, the phase sensitivity approaches a constant. As a result, the superresoluion disappears and the output fringes approach the SNL distribution quickly. Only in the low-photon-number regime ($n \le 6$), is it possible to beat the SNL. In the experiment, the effective visibilities of the measured fringes are $0.94 \pm 0.02$ and $0.97 \pm 0.03$ for $n=3$ and 4, respectively. They are greater than the corresponding thresholds ($0.83$ and $0.93$) to beat the SNL. As shown in Table~\ref{tab:table}, all the measuring phase sensitivities beat the SNLs. More details can be found in the Supplemental Material \cite{SUPPLEMENTALMATERIAL}.

In conclusion, we have experimentally demonstrated the generalized HOM effect in QFT interferometers and observed phase supersensitivies in the multimode MZIs with two, three, and four photons. Our simple QFT devices may be used to realize other QFT-based applications, such as quantum-enhanced multiphase estimation \cite{humphreys2013,ciampini2016,*szczykulska2016}, entanglement generation and transformation \cite{lim2005b}, sorting quantum systems efficiently \cite{ionicioiu2016}, nonmonotonic quantum-to-classical transitions \cite{tichy2011,*ra2013}, and simulations of geometric phase \cite{laing2012}. Additionally, the scheme using both path and polarization modes are also suited to optical waveguide systems for simplifying the construction of QFT interferometers.

{\em Acknowledgments---} This work was supported by the National Natural Science Foundation of China, the Chinese Academy of Sciences, and the National Fundamental Research Program. JPD would like to acknowledge support from the US National Science Foundation. PPR is funded by an ARC Future Fellowship (project FT160100397). All the authors would like to acknowledge Chenglong You, Sushovit Adhikari, Jonathan Olson, and Tim Byrnes for helpful discussions.

\providecommand{\noopsort}[1]{}\providecommand{\singleletter}[1]{#1}%

\section{SUPPLEMENTAL MATERIAL}
\renewcommand{\thefigure}{S\arabic{figure}}
\setcounter{figure}{0}
\renewcommand{\theequation}{S.\arabic{equation}}
\setcounter{equation}{0}
\renewcommand{\thesection}{S.\Roman{section}}
\setcounter{section}{0}

\section{Single-photon source characterization}

The ultraviolet pumping laser which we use has central wavelength of 390 nm, pulse duration of 100 fs, repetition rate of 80 MHz. The pulses pump collinear Type-II cut BBO crystals and generate PDC photon pairs. In order to minimize higher order noise, we deliberately lower the average pump power to 650 mW while maintaining a feasible brightness. We estimate a single pair generation efficiency $\sim$0.02, and a collection and detection efficiency $\sim$0.26. 

\begin{figure*}
	\includegraphics[width=6.0 in]{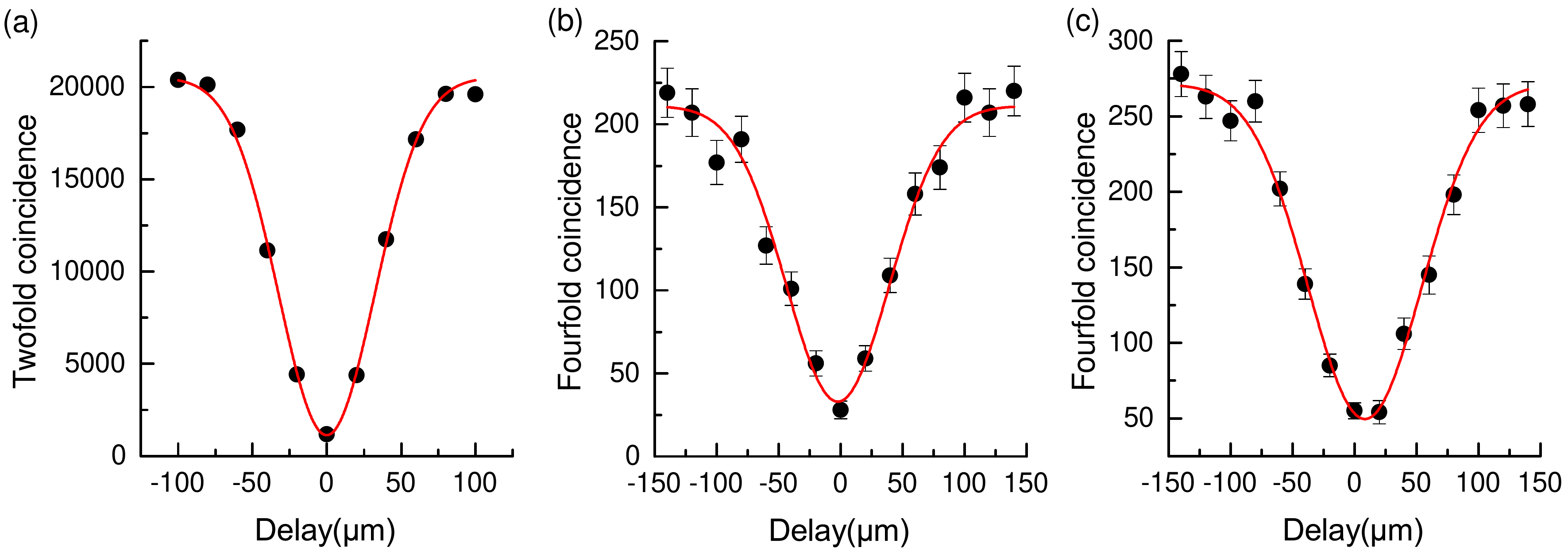}
	\renewcommand\thefigure{S\arabic{figure}}
	\caption{\label{fig:hom} (color online). Observation of HOM interference. The interference fringes visibilities are (a) $0.945 \pm 0.002$ between photons 2 and 3, (b) $0.84 \pm 0.03$ between photons 1 and 3, and (c) $0.82 \pm 0.03$ between photons 3 and 4. Error bars are one standard deviation due to propagated Poissonian statistics.}
\end{figure*}

Accurate spatial and temporal mode overlap are necessary to make indistinguishable independent photons. To achieve good spatial and temporal overlap, all photons are spectrally filtered and detected by fiber-coupled single-photon detectors. To find the optimal indistinguishability, we scanned the delay lines $\Delta {d_1}$ - $\Delta {d_3}$ and observed HOM interference fringes, as shown in Fig.~\ref{fig:hom}.

Imperfection of optical elements is another main cause for experimental error. For example, the splitting ratio tolerance ($\pm3\%$) of beam splitter will make the interferometer deviate from the quantum Fourier transform slightly. Total efficiency of the 3 (4) modes optical interferometer is $\sim$0.87 ($\sim$0.94). As to interference stability, the interferometers which we build are compact and stable enough ($>$ 1 day) to finish the measurement.

\section{Optical interferometer construction}

\begin{figure*}
	\includegraphics[width=4.0 in]{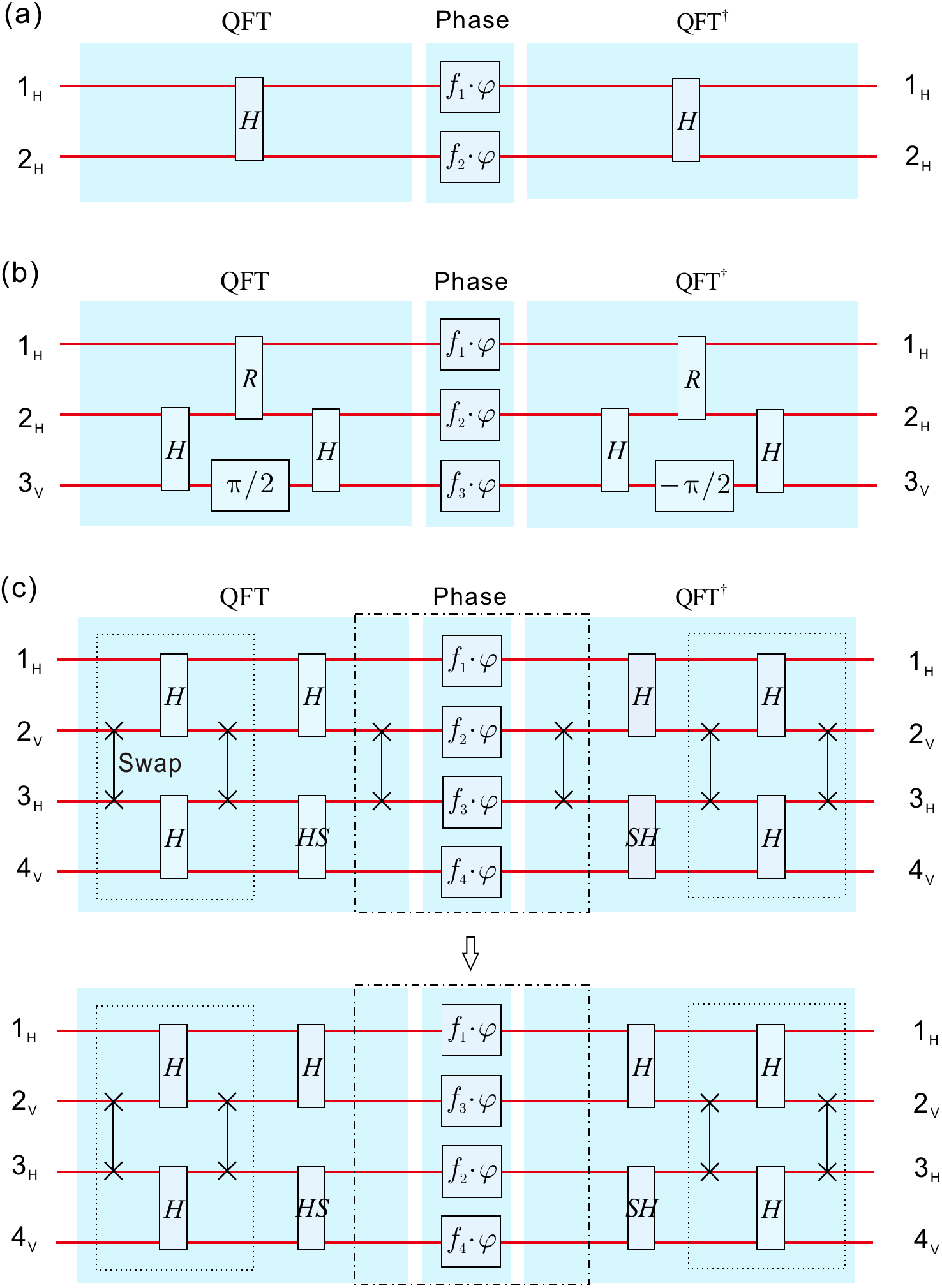}
	\renewcommand\thefigure{S\arabic{figure}}
	\caption{\label{fig:qftdecomposition} (color online). Quantum-circuit representation of the experimental setup in Fig.2. (a) $n = 2$. (b) $n = 3$. (c) $n = 4$.}
\end{figure*}

For $n=2$, the QFT is equivalent to the Hadamard matrix
\[H = \frac{1}{{\sqrt 2 }}\left( {\begin{array}{*{20}{c}}
	1&1\\
	1&{ - 1}
	\end{array}} \right),\]
which can be realized by an NBS with two path modes [see Fig.~\ref{fig:qftdecomposition}(a)].

For $n=3$, three modes are represented by (${\left| H \right\rangle _1}$, ${\left| V \right\rangle _1}$, ${\left| H \right\rangle _2}$), where subscript 1 and 2 refer to path modes. The QFT can be realized by a PDBS with reflectivity 1/3 for horizontal polarization and perfect reflectivity for vertical polarization [see Fig.~\ref{fig:qftdecomposition}(b)]. The function of the PDBS could be written by
\[R \oplus {I_3} = \frac{1}{{\sqrt 3 }}\left( {\begin{array}{*{20}{c}}
	1&{\sqrt 2 }\\
	{\sqrt 2 }&{ - 1}
	\end{array}} \right) \oplus {I_3}.\]

For $n=4$, four modes are represented by (${\left| H \right\rangle _1}$, ${\left| V \right\rangle _1}$, ${\left| H \right\rangle _2}$, ${\left| V \right\rangle _2}$). The QFT interferometers can be realized by a NBS followed by a polarizing Hadamard matrix $ H $ in one path, which can be realized by a HWP, and a polarizing transform
\[HS = \frac{1}{{\sqrt 2 }}\left( {\begin{array}{*{20}{c}}
	1&1\\
	1&{ - 1}
	\end{array}} \right)\left( {\begin{array}{*{20}{c}}
	1&0\\
	0&i
	\end{array}} \right) = \frac{1}{{\sqrt 2 }}\left( {\begin{array}{*{20}{c}}
	1&i\\
	1&{ - i}
	\end{array}} \right)\]
in another path, which can be realized by a QWP and a HWP [Fig. ~\ref{fig:qftdecomposition}(c)].

\section{Simplifying QFT interferometers using path and polarization modes}
Given a $d$-mode non-polarizing QFT interferometer ${F_{d \times d}}$, we can expand it to ${F_{2d \times 2d}}$ as,
\[{F_{2d \times 2d}} = P[ \oplus _{k = 0}^{d - 1}H(k\pi /d)][{F_{d \times d}} \oplus {F_{d \times d}}]P,\]
where $H(\theta ) = \frac{1}{{\sqrt 2 }}\left( {\begin{array}{*{20}{c}}
	1&{{e^{i\theta }}}\\
	1&{ - {e^{i\theta }}}
	\end{array}} \right)$ is the product of Hadamard matrix and phase shift matrix, and $P$ is the $2d \times 2d$ permutation matrix that maps the column vector $\vec v  = {({H_1},{V_1},{H_2},{V_2} \cdots ,{H_d},{V_d})^T}$ into
\[P\vec v  = {({H_1},{H_2} \cdots ,{H_d},{V_1},{V_2}, \cdots {V_d})^T},\]
that is, it shuffles components of $\vec v $ such that horizontal polarizations appear on the first half and vertical polarizations on the latter part.

The scheme is essentially equivalent to the first stage of the fast Fourier transform, i.e., the famous factorization discovered by Cooley and Tukey \cite{cooley1965} (also by Gauss \cite{strang1993} in 1805). The general method \cite{reck1994,*clements2016} requires $O({m^2})$ beam splitters to realize an arbitrary $m$-dimensional unitary transformation, which was recently adopted to realize a six-mode QFT interferometer in a planar light-wave circuit \cite{carolan2015}. With the help of the aforementioned technique, the number of beam splitters can be reduced by 75\%. Thanks to the special form of the QFT, more efficient methods \cite{torma1996,*barak2007,*tabia2016} only requiring $O(m\log (m))$ beam splitters have been proposed and used to construct an eight-mode QFT interferometer in a complex 3D waveguide circuit \cite{crespi2016}. In this case, the number of beam splitters can still be reduced by over 50\%.

\section{Classical probability distributions of the QFT interferometers}

\begin{figure*}
	\includegraphics[width=6.0 in]{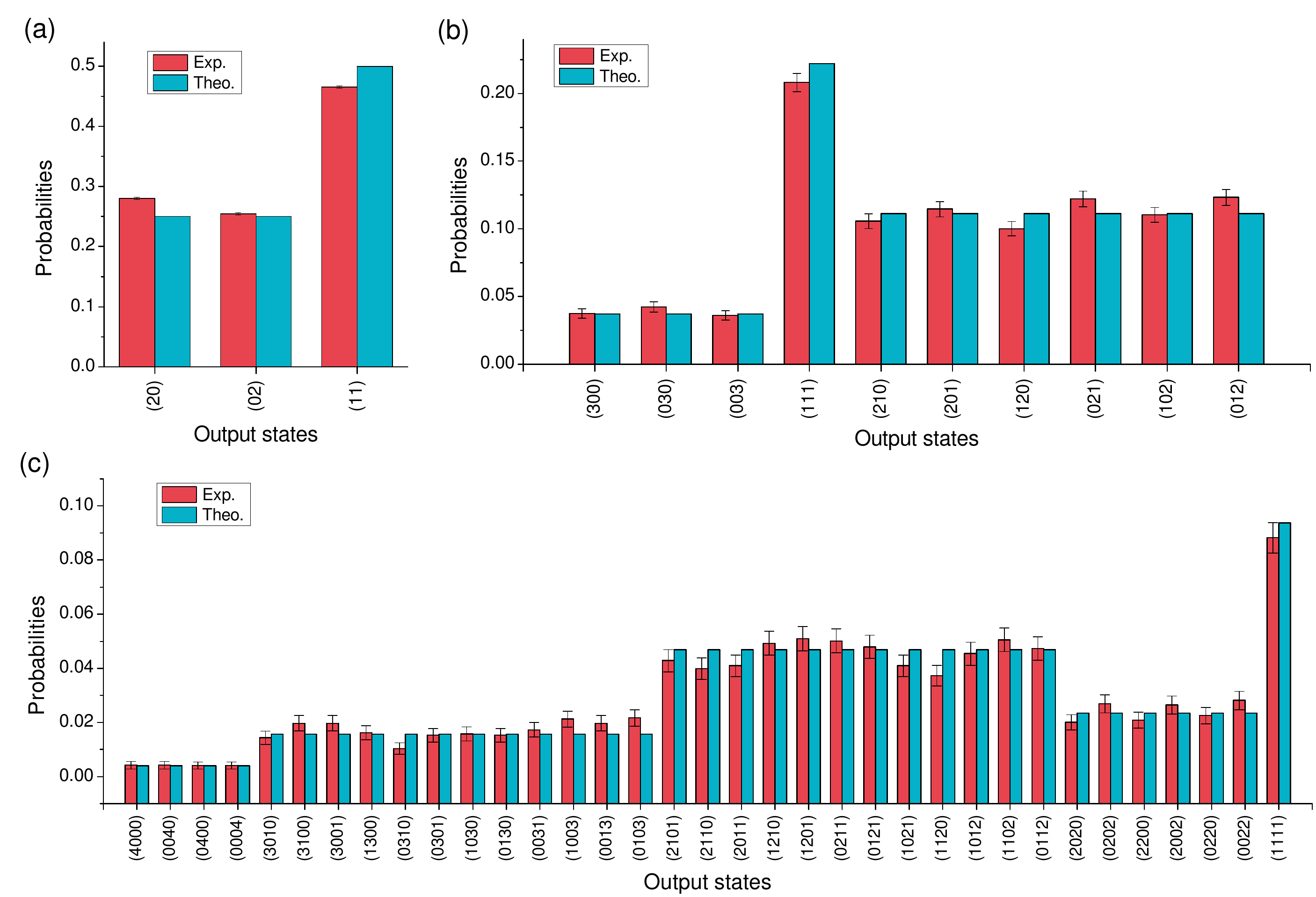}
	\renewcommand\thefigure{S\arabic{figure}}
	\caption{\label{fig:qftnon} (color online). Experimental results of the QFT interferometers with distinguishable single-photon inputs. (a) $n = 2$. (b) $n = 3$. (c) $n = 4$. We obtained fidelities as (a) $ 0.999 \pm 0.002$, (b) $ 0.999 \pm 0.001$ and (c) $ 0.998 \pm 0.001$. Error bars are one standard deviation due to propagated Poissonian statistics.}
\end{figure*}

For comparison, we measure the output probability distributions in the QFT interferometers with distinguishable single-photon inputs (which can be modelled by inserting identical coherent-state inputs). The results are shown in Fig.~\ref{fig:qftnon}. The violations of different input single photons are $\upsilon _2^{\mbox{d}} = 0.47 \pm 0.01$, $\upsilon _3^{\mbox{d}} = 0.68 \pm 0.08$, $\upsilon _4^{\mbox{d}} = 0.75 \pm 0.14$. The results show that the generalized HOM effect is distinctly different with indistinguishable or distinguishable photons.

\section{Normalized linear phase versus delta phase}
In the main text, we have explained why we cannot use the QFT interferometers with linear phase distribution to beat the SNL. The root of the problem is the utilization of multiple phases $\varphi$. In order to compare different strategies fairly, we adopted the normalization condition $\sum\nolimits_i {{f_i} = 1} $ \cite{olson2016linear} and redrew the fringes of output probability as function of phase shift $\varphi$. As shown in Fig.~\ref{fig:compare}, we can conclude that the fringe with higher phase sensitivity also has steeper slope near the origin.

\section{Phase sensitivity calculation}

The experimental data can be fitted by the fringes as
\begin{equation}
\mbox{Counts} = \frac{{V(2p - 1) + 1}}{2} \cdot A,
\end{equation}
where $p$ is the theoretical probability \cite{olson2016linear}, and $A$ and $V$ are fitted parameters denoting the largest count rate and effective visibility, respectively. According to the Cramer-Rao bound, the phase sensitivities can be determined as $\Delta\varphi = 1/\sqrt {\mbox{FI}} $, where FI denotes the Fisher information,
\begin{equation}
\label{eq:FI}
\mbox{FI} = \frac{{4{V^2}{{\left| {\partial p/\partial \varphi } \right|}^2}}}{{1 - {V^2}{{(2p - 1)}^2}}}.
\end{equation}
The error of FI can be deduced from the error of $V$ via Eq.~\ref{eq:FI} and error propagation formula.

\begin{figure*}
	\includegraphics[width=4.0 in]{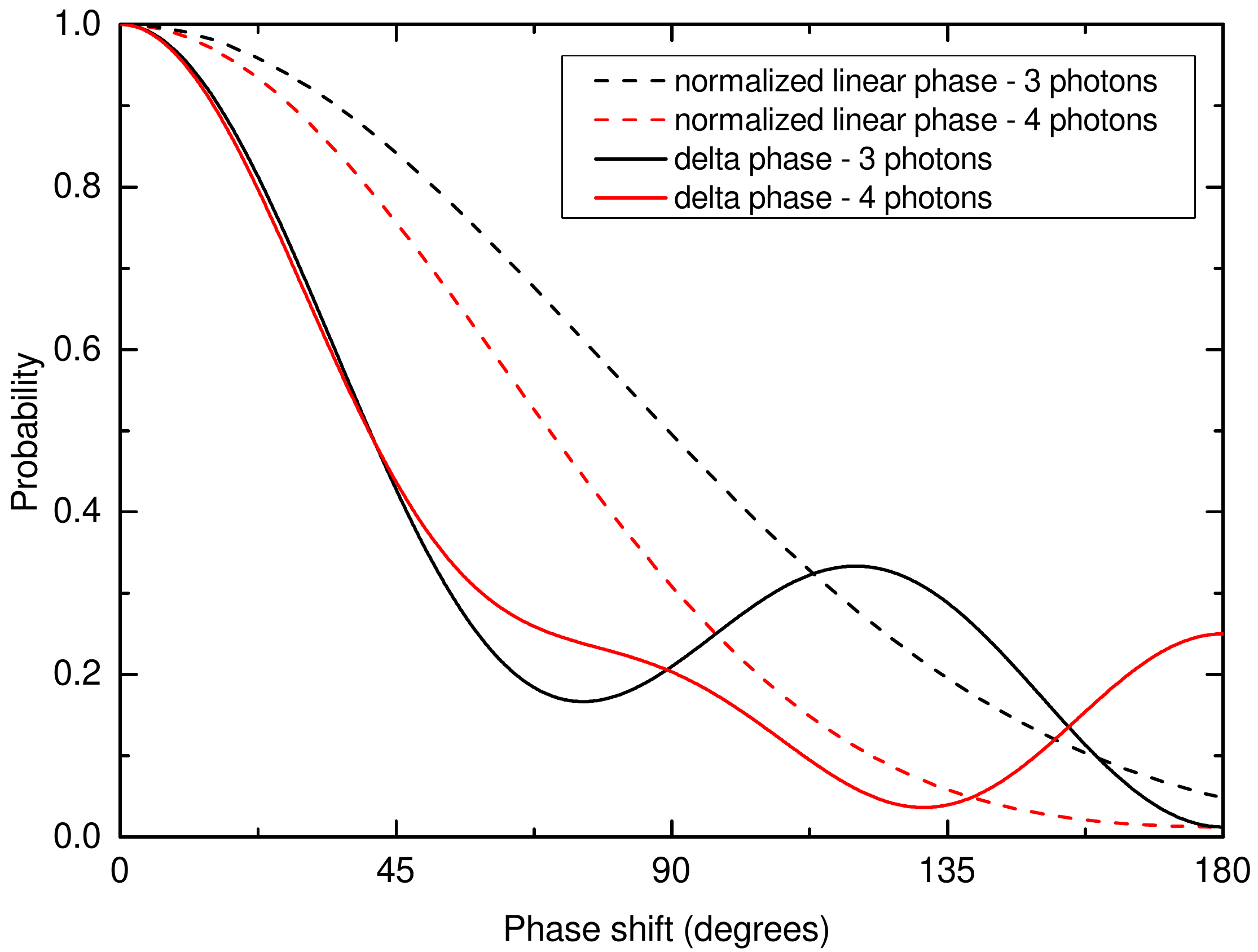}
	\renewcommand\thefigure{S\arabic{figure}}
	\caption{\label{fig:compare} (color online). Normalized linear phase versus delta phase.}
\end{figure*}

For the delta-phase scheme, letting $\mbox{FI} > n$, we can obtain the thresholds of the effective visibility to beat the SNL. The thresholds are $0.708$, $0.826$ and $0.923$ for $n = 2$, 3 and 4, respectively. In Eq.~\ref{eq:FI}, $p$ at the origin and $V$ are independent of the phase $\varphi$. It is then easy to see that the schemes with the largest FI (and consequently the smallest phase uncertainty) will have the steepest (largest) slope of $p(\varphi)$ near the origin. That this is so can be seen clearly in the data plots in Fig.~\ref{fig:compare}.

\providecommand{\noopsort}[1]{}\providecommand{\singleletter}[1]{#1}%

\end{document}